\documentclass{article}
\usepackage{amsmath}
\usepackage{graphicx}

\begin{document}

\begin{center}
{\bf \Large Space Dependence of Entangled States and}
\end{center}
{}

{}

\begin{center}
{\bf \Large Franson-type EPR Experiments\bigskip }

A.A. Baranov, A.N. Pechen and I.V. Volovich{\Large \bigskip }

Steklov Mathematical Institute

Russian Academy of Science

Gubkin St. 8, GSP1, 119991, Moscow

Russia

{\it email: volovich@mi.ras.ru}

{\Large \bigskip \bigskip }

\textbf{Abstract}
\end{center}

{\small We analyze some aspects of recently performed Franson-type
experiments with entangled photon pairs aimed to test Bell's inequalities.
We point out that quantum theory leads to the coincidence rate between
detectors which includes in fact a dependence on the distance. We study this
dependence and obtain that for large distances the correlation function
vanishes. Therefore with taking into account the space parts of wave
functions of photons for large distances quantum mechanical predictions are
consistent with Bell's inequalities. We propose an experimental study of
space dependence of correlation functions in \ Bell-type experiments.}%
\bigskip \bigskip

\section {Introduction}

Violation of Bell's inequalities is the subject of numerous
theoretical and experimental investigations $[1,2]$. Most of EPR-type
experiments used optical photons and requires the use of polarizers. But
Franson $[3,4]$ and others researchers $[5-7$, $12]$ pointed out that we
have the possibility of experimental testing of Bell's inequalities by
alternative ways. One can get a violation of Bell's inequalities as a result
of interference between the probability amplitudes for a pair of photons
emitted at various times by an excited atom. The quantum-mechanical
uncertainly in the position of a particle or the time of its emission is
shown to produce observable effects that are inconsistent with any local
hidden-variable theory if one neglects the space part of the wave function.

This experiment is based on optical interference. However the
transmission of photons in all the types of EPR-experiment is a real
physical process in space and time. Therefore it is important to study the
spacetime dependence of correlation functions. The importance of
consideration of the spacetime dependence in Bell's type computations in
quantum theory was pointed out in $[9-11]$. In this paper we consider the
spacetime dependence of correlation functions in Franson-type experiments
for entangled states.

Consider a pair of spin one-half particles in the singlet spin state.
If we neglect the spacetime dependence of the wave function the quantum
mechanical correlation of two spins is
\begin{equation}
E_{spin}(a,b)=\left\langle \psi _{spin}|\sigma \cdot a\otimes
\sigma \cdot b|\psi _{spin}\right\rangle =-a\cdot b
\end{equation}
Here $a$ and $b$ are two unit vectors in space and $\ \sigma =(\sigma
_{1},\sigma _{2},\sigma _{3})$ are the Pauli matrices. Bell's theorem states
that \ this can not be represented in the form of classical correlation
function $P(a,b)$. Bell-Clauser-Horn-Shimony-Holt (CHSH) form of Bell's
inequality of two stochastic processes is
\begin{equation}
|P(a,b)-P(a,b^{\prime })+P(a^{\prime },b)+P(a^{\prime },b^{\prime
})|\leq 2
\end{equation}

From the other hand if $ab=a^{\prime }b=a^{\prime }b^{\prime }=-ab^{\prime }=%
\sqrt{2}/2$ then
$$|E_{spin}(a,b)-E_{spin}(a,b^{\prime })+E_{spin}(a^{\prime
},b)+E_{spin}(a^{\prime },b^{\prime })|=2\sqrt{2}
$$
and Bell's inequality is violated.

Now we suppose that detectors are located in two separated regions $%
O_{1\text{ }}$and $O_{2}$ respectively and perform localized observations.
In this case the quantum correlation is
\begin{equation}
E(a,O_{1},b,O_{2})=\left\langle \psi |\sigma \cdot aP_{O_{1}}\otimes \sigma
\cdot bP_{O_{2}}|\psi \right\rangle
\end{equation}
Here $P_{O}$ is the projection onto the region $O.$ If the wave function has
a special form of the product of the spin part and the space part,\ $\psi
=\psi _{spin}\phi (r_{1},r_{2})$ then
\begin{equation}
\label{1}
E(a,O_{1},b,O_{2})=g(O_{1},O_{2})E_{spin}(a,b)
\end{equation}
where the function
$$g(O_{1},O_{2})=\int_{O_{1}\times O_{2}}|\phi (r_{1},r_{2})|^{2}dr_{1}dr_{2}
$$
describes correlation of particles in space. The factor $g(O_{1},O_{2})$ is
important. We obtain that Bell's inequalities can be violated only if
\begin{center}
$ g(O_{1},O_{2})>1/\sqrt{2}$
\end{center}

The factor $g(O_{1},O_{2})$ deserves an experimental studying. It was shown
\ in $[11]$ that at large distances any quantum state becomes disentangled.

That's why in the present paper we analyze the Franson test of Bell's
inequalities in a particularly simple configuration consisting of two
separated interferometers driven by photons from the source. Our goal is to
investigate the space dependence of quantum correlation function in this
case. We will try to connect the value of coincidence rate $R_{0}$ and a
possibility of violation of Bell's inequality. The decrease of value of $%
R_{0}$ with the increase of distances between detectors and the source of
photons is a crucial point for violation of Bell's inequalities.

The interference experiment can be described in the following way
(from work of Franson $[3]$). At time $t=0$ an atom is assumed to be excited
into the upper state $\psi _{1}$,\ which has a relatively long lifetime\ $%
\tau _{1}$. After emission of a photon $\gamma _{1}$ with wavelength $%
\lambda _{1}$ the atom will be in the intermediate state $\psi _{2}$,\ which
has a relatively short lifetime $\tau _{2}\ll \tau _{1}$. Thus a second
photon $\gamma _{2}$ with wavelength $\lambda _{2}$ will be emitted very
soon after $\gamma _{1}$ and a coincidence counting experiment would show a
very narrow peak with a width $\sim \tau _{2}$. Photons $\gamma _{1}$and $%
\gamma _{2}$ are collimated by lenses L$_{1}$ and L$_{2}$ into beams which
propagate toward distant detectors D$_{1}$ and D$_{2}$, respectively. The
coincidence counting rate will simply show a narrow peak indicating that $%
\gamma _{1}$and $\gamma _{2}$ were emitted at times which were the same to
within a small uncertainty $\sim \tau _{2}.$

The quantum-mechanical description of this process is highly nonlocal
in space since the time at which either photon was emitted was initially
uncertain over a much larger time interval $\sim \tau _{1}$. As a result,
the two photons must initially be described by wave packets in which their
time of emission and thus their position is relatively uncertain. The
detection of one of the photons, say $\gamma _{1}$, immediately determines
the time of emission of the other photon and thus its position to within a
much smaller uncertainty, which must be reflected by a nonlocal change in
the wave function describing the other photon. This nonlocal reduction of
the wave function is analogous to that which occurs in the polarization
measurements of Bell's original theorem. The relative phases of the photons
play the role of polarization angles in Bell's type experiments.

\section { Correlation function}

Let us investigate the dependence on the distance in the coincidence
experiment with the half-silvered mirrors in place. To calculate the
coincidence rates one introduced $[3]$ the field operator
$$
\psi (r)=\int e^{ik\cdot r}a(k)dk
$$
where $a(k)$ annihilates a particle with momentum $k$ and $r$ is a distance
from the source.  The time dependence of the field operator is given
by
$$
\psi (r,t)=e^{itH/\hbar }\psi (r)e^{-itH/\hbar }
$$
where $H$ is the Hamiltonian of the system.

The\ field at detector D$_{1}$ is given by
$$
\psi (r_{1},t)=\frac{1}{2}\psi _{0}(r_{1},t)+\frac{1}{2}e^{i\phi _{1}}\psi
_{0}(r_{1},t-\Delta T)
$$
Here $\psi _{0}(r,t)$ is the field operator with the half-silvered mirrors
removed and $\Delta T$ is the difference between in the transit times via
the longer and shorter paths, $\phi _{1},\phi _{2}$ (see below) are phase
shifts into the two beams.

The field at the detector D$_{2}$ is
$$\psi (r_{2},t)=\frac{1}{2}\psi _{0}(r_{2},t)+\frac{1}{2}e^{i\phi _{2}}\psi
_{0}(r_{2},t-\Delta T)
$$

The coincidence rate $R_{c}$ between D$_{1}$ and D$_{2}$ with the
mirrors inserted can now be calculated from
\begin{center}
$R_{c}=\eta _{1}\eta _{2}$ $\left\langle 0\mid \psi ^{\dagger }(r_{1},t)\psi
^{\dagger }(r_{2},t)\psi (r_{2},t)\psi (r_{1},t)\mid 0\right\rangle $
\end{center}
where $\eta _{1}$ and $\eta _{2}$ are the detection efficiencies of D$_{1}$
and D$_{2}$ and $|0\rangle $ \ is the vacuum state. One can compute $R_{c}$
to be of the form
\begin{equation}
 R_{c}=\frac{1}{4}R_{0}\cos ^{2}\left( \phi
_{1}^{\prime }-\phi _{2}^{\prime }\right)
\end{equation}
where
$$R_{0}=\left\langle 0\mid \psi _{0}^{\dagger }(r_{1},t)\psi _{0}^{\dagger
}(r_{2},t)\psi _{0}(r_{2},t)\psi _{0}(r_{1},t)\mid 0\right\rangle 
$$
is the coincidence rate with the half-silvered mirrors removed.

Let us study the dependence $R_{0}$ on the distances $r_{1}$ and $%
r_{2}$ in a simple model.

\section {Model}

Let us consider the quantum model with the Hamiltonian
$$H=H_{0}+V=\int dk\omega (k)a^{\dagger }(k)a(k)+\int dk(\overset{-}{f}%
(k)a(k)+f(k)a^{\dagger }(k))
$$
Here $f(k)$ is a formfactor (test function) and $\omega (k)$ is a dispersion
low. The field operators (in coordinate representation) are
$$\psi _{0}(r)=\int dpa(p)e^{ipr}
$$

In the Heisenberg representation the states remain constant while the
operators evolve in time so that
\begin{equation}
\label{2}
\psi _{0}(r,t)=e^{itH}\psi _{0}(r)e^{-itH}=\int
dpe^{ipr}f_{p}(t)
\end{equation}
Here
$$f_{p}(t)=e^{itH}a(p)e^{-itH}=a(p)e^{-i\omega (p)t}+\frac{f(p)}{\omega (p)}%
(e^{-i\omega (p)t}-1)
$$

Our goal is to study the quantity
\begin{center}
$R_{0}=\left\langle 0\mid \psi _{0}^{\dagger }(r_{1},t)\psi
_{0}^{\dagger }(r_{2},t)\psi _{0}(r_{2},t)\psi _{0}(r_{1},t)\mid
0\right\rangle $
\end{center}
Using~(\ref{2}) one immediately gets
$$R_{0}=\int dp_{1}dp_{2}dp_{1}^{\prime }dp_{2}^{\prime
}e^{-ip_{1}r_{1}-ip_{2}r_{2}+ip_{1}^{\prime }r_{1}+ip_{2}^{\prime
}r_{2}}\left\langle 0\mid f_{p_{1}}^{\dagger }(t)f_{p_{2}}^{\dagger
}(t)f_{p_{2}^{\prime }}(t)f_{p_{1}}(t)\mid 0\right\rangle
$$
$$=\int dp_{1}dp_{2}dp_{1}^{\prime }dp_{2}^{\prime
}e^{-ip_{1}r_{1}-ip_{2}r_{2}+ip_{1}^{\prime }r_{1}+ip_{2}^{\prime
}r_{2}}\frac{\overset{}{\overline{f}(p_{1})}}{\omega (p_{1})}%
(e^{i\omega (p_{1})t}-1)\frac{\overset{}{\overline{f}(p_{2})}}{\omega
(p_{2})}(e^{i\omega (p_{2})t}-1) 
$$
$$\times \frac{\overset{}{f(p_{1}^{\prime })}}{\omega (p_{1}^{\prime })}%
(e^{-i\omega (p_{1}^{\prime })t}-1)\frac{\overset{}{f(p_{2}^{\prime })%
}}{\omega (p_{2}^{\prime })}(e^{-i\omega (p_{2}^{\prime })t}-1)=\left|
\varphi (r_{1},t)\right| ^{2}\left| \varphi (r_{2},t)\right| ^{2}
$$
Here
$$\varphi (r,t)=\int dpe^{ipr}\frac{f(p)}{\omega (p)}(e^{-i\omega (p)t}-1)
$$

Let us consider two cases. The first case the formfactor $f(p)$ has
an exponential form and the second case the formfactor has the form of a
simple cut-off.

In order to investigate the behavior of $\varphi (r,t)$ as $%
|r|\rightarrow \infty $ let us consider some concrete examples of formfactor
and dispersion low. Suppose that our Hamiltonian describes massless
particles, i.e. $\omega (p)=|p|$, $p\in R^{3}.$ We will consider two types
of formfactor - the step-function and a gaussian formfactor. In both cases
formfactor depends only on $|p|$. Therefore for $\varphi (r,t)$ one
has
$$\varphi (r,t)=\frac{2\pi }{i|r|}\int_{0}^{\infty
}d|p|f(|p|)(e^{-i|p|t}-1)(e^{i|p||r|}-e^{-i|p||r|})
$$
$$
=\frac{2\pi }{i|r|}%
(I(t-|r|)-I(t+|r|)+I(|r|)-I(-|r|))
$$
Here we denoted
\begin{center}
$I(\alpha )=\int_{0}^{\infty }dxf(x)e^{-i\alpha x}$
\end{center}

Now we consider the case $f(|p|)=\theta (|p|-A)$ where $\theta $ is the step
function for some positive constant $A$. We have
\begin{center}
$I(\alpha )=i\frac{e^{-i\alpha A}-1}{\alpha }$
\end{center}

This means that $\left| \varphi (r,t)\right| ^{2}$ vanishes for large $%
|r|$ at least as $1/|r|^2$.

For the gaussian formfactor $f(|p|)=e^{-|p|^{2}/A}$ the asymptotics 
has the same form $\left| \varphi (r,t)\right| ^{2}\sim 1/\left|
r\right| ^{2}$.

This means that $R_{0}$ polynomially decreases at large distances
between detectors.

Moreover one can slightly modify the gaussian formfactor in some
neighborhood of $p=0$ in such a way that $R_{0}$ will decrease faster then
any polynomial.
To obtain such a behaviour just take $f(x)$ to be a test function
with a compact support on the positive semiaxis. \bigskip \bigskip

\section { Conclusions}\bigskip

In this paper the space dependence of correlation function in
Franson-type experiments is discussed. The role of the spacetime dependence
of the correlation function in Bell-type experiments was considered earlier
in $[9-11]$. The Franson-type experiment that is taken as a basis of this
paper is a new variant of EPR-type experiments (according to $[3-7]$ and
others researchers) and the form of $R_{c}$ in Eq.~(5) is identical to
that obtained in earlier experiments based upon Bell theorem where $\phi
_{1}^{\prime }$ and $\phi _{2}^{\prime }$ correspond instead to the
orientation of distant polarizers $[3]$.

In our case the space dependent coefficient $R_{0}$ can play the role
of the function $g(O_{1},O_{2})$ in Eq (4). Although in work of W. Tittel et
al.$[6]$ correlation coefficient $E$ is independent of $R_{0}$ ($R_{0}$ is
canceled out in process of calculation of $E$) but Larsson et al. $[8]$
suggest that Franson-type experiments can not demonstrate violation of
Bell's inequality. Authors notes that only half of the events will belong to
coincidence class and if all that are taken into account standard Bell
inequalities are not violated. They suggested the quantum model predicted
probability that equal $1/2$ of the probability of photon coincidence from $%
[6]$. Hence $1/2$ is a coefficient in left part of Bell inequality and
that's why it is not violated. In $[12]$ also one showed the dependence of
probability and visibility value on number of registered events.

In $[6,7]$ one proposed to use the visibility $V$ as a
coefficient before the function of cosine of phase differences. If $V$ $\leq
$ $1/\sqrt{2}$ then there is no violation of the Bell's inequalities and
therefore there is no violation of locality in the corresponding state of
wave function. Authors notes that visibility decreased compared to short
distance experiments but without obvious visibility dependence on distance.
In $[6]$ the maximal visibility is a function of counts $(C,$ $A)$ were $C$
is the number of detected coincidences and $A$ is the number of accidental
ones.

Therefore we need a careful analysis of number of discarded events
because this is a crucial point for detecting a possible violation of Bell's
inequality. But in real EPR-type experiments this number depends on a
distance between two separated regions $O_{1}$ and $O_{2}$ where detectors
are located. In our case the increasing of the distance between detectors
and the source of photons leads to the decreasing of the coefficient $R_{0}$ that
plays the role of space part of correlation function. We suggest that there
is a connection between the space depended function $g(O_1,O_2)$ and the 
coefficient $R_{0}$.

When $g(O_1,O_2)$ becomes smaller than $1/\sqrt{2}$ Bell's inequality is not
violated and therefore for large distances quantum mechanical predictions
are consistent with Bell's inequalities. That's why it is possible that the
value $R_{0}$ plays a key role in this line of Franson-type experiments and
really we must perform a study of the dependence of $R_{0}$ on the distance.
This question requires further theoretical and experimental investigations.\bigskip \bigskip

\section{Acknowledgments}

This work is partially supported by  RFFI 02-01-01084, by INTAS
99-00545, the grant for leading scientific schools RFFI 00-15-96073
and also by the INTAS 01/1-200 for A.P.


\bigskip \bigskip

\begin{thebibliography}{99}

\bibitem{b1} J. S. Bell, Physics 1, 195 (1964)

\bibitem{w1} R. F. Werner, M. M. Wolf, \textit{Bell inequalities and Entanglement, }%
quant-ph/0107093
\bibitem{f1} J. D. Franson, Phys.Rev. Lett., 62, 2205 (1989)
\bibitem{f2} J. D. Franson and K. A. Potocki, Phys. Rev. A 37, 2511
(1988)
\bibitem{t1} W. Tittel, J. Brendel, H. Zbinden, N. Gisin, \textit{%
Violation \ of Bell inequalities by photons more than 10 km apart,}
quant-ph/9806043
\bibitem{t2} W. Tittel, J. Brendel, N. Gisin, H. Zbinden, \textit{Long-distance
Bell-type tests using energy-time entangled photons, }quant-ph/9809025 v1
\bibitem{b2} J. Brendel, E. Mohler and W. Martienssen, Europhys. Lett.
20 (7), 575 (1992)
\bibitem{l1} J. Larsson, S. Aerts and M. Zukowski, \textit{%
Two-photon Franson-type interference experiments are not tests of local
realism,} quant-ph/9812053
\bibitem{v1} I. V. Volovich, \textit{Bell's theorem and locality in
space,} quant-ph/0012010
\bibitem{v2} I. Volovich, \textit{Quantum cryptography in space and Bell's theorem.}%
In \textit{Foundation of probability and physics.} pp. 364-372. Ed. A.
Khrennikov, World Scientific, 2001
\bibitem{v3} I. V. Volovich, \textit{Towards quantum information
theory in space and time,} quant-ph/0203030
\bibitem{b3} J. Brendel, E. Mohler and W. Martienssen, Phys.Rev. Lett., 66 (9),
1142 (1991)
\end{thebibliography}
\end{document}